\documentclass[aps,twocolumn,pra,showpacs,floatfix]{revtex4}
\usepackage{epsfig}
\usepackage{graphicx}
\usepackage{dcolumn}
\usepackage{amsmath}

\begin{document}

\title{Dynamic polarizabilities and magic wavelengths for dysprosium}
\author{V. A. Dzuba and V. V. Flambaum}
\affiliation{School of Physics, University of New South Wales,
Sydney 2052, Australia}
\author{Benjamin L. Lev}
\affiliation{Department of Physics, University of Illinois at
  Urbana-Chapaign, Urbana, Illinois 61801-3080, USA}
\date{\today}

\begin{abstract}

We theoretically study dynamic scalar polarizabilities of the ground
and select long-lived excited states of dysprosium, a highly magnetic
atom recently laser cooled and trapped. We demonstrate that there are
a set of magic wavelengths of the unpolarized lattice laser 
field for each pair of states which includes the ground state and one of these
excited states.  At these wavelengths, the energy shift due to laser
field is the 
same for both states, which can be useful for resolved sideband
cooling on narrow transitions and precision spectroscopy.  We present
an analytical formula which, near resonances, allows for the
determination of approximate values of the magic wavelengths without
calculating the dynamic polarizabilities of the excited states.   
 
\end{abstract}


\maketitle

\section{Introduction}

The dysprosium atom has many unique features which makes it useful for
studying fundamental problems of modern physics. This is a heavy atom
which has many stable Bose and Fermi isotopes (from $A=156$ to $A=164$) and a pair of
almost degenerate states of opposite parity at
E=19798~cm$^{-1}$. These features were used in study of the parity
non-conservation
(PNC)~\cite{Dzuba86,Dzuba94,Budker94,Budker97,Dzuba10} and 
possible time-variation of the fine structure
constant~\cite{Dzuba99a,Dzuba99b,Dzuba03,Budker04,Budker07,Dzuba08,Budker07a}.  

Fermionic Dy has the largest magnetic moment among all atoms, and only
Tb is as magnetic as bosonic Dy. This opens 
important opportunities in studying strongly correlated matter when
gases of Dy atoms is cooled to ultracold
temperatures~\cite{Lev}. Recent progress in Doppler and sub-Doppler
cooling is an important step in this
direction~\cite{Budker08,Lev,Leefer,Youn2010a,Youn2010b}. In addition
to narrow-line magneto-optical trapping (MOT)~\cite{Berglund:2008},
further cooling on narrow optical transitions might be possible using
resolved-sideband cooling~\cite{Wineland89,Katori03}.  

In this method, vibrational states of the atom may coupled such that
successive photon absorption and spontaneous emission cycles reduce
the vibrational quanta by one, until the atoms are in the motional
ground state of their optical potential~\cite{Wineland89}.  It is
important that this resolved-sideband cooling is performed at {\em
  magic} wavelength of the 
laser lattice field~\cite{Katori08,Riehle07}.  At this wavelength, the
energy (AC Stark) shift due to laser field is the same for both states
used in the cooling.  This results in a trap potential that is the
same for both states, and optical transitions between vibrational
states can be well resolved.  This allows spectral selection of
cooling transitions---those which remove one vibrational
quanta---without contamination by heating transitions which add
vibrational quanta.  Other benefits to optical trapping at magic
wavelengths include enhanced precision spectroscopy and longer-lived
quantum memory for quantum information processing
(QIP)~\cite{Katori08}. 

In this paper we calculate dynamic polarizabilities of the ground and
three long-lived excited states of Dy and present a number of
magic wavelengths for the transitions between them.  We also present
an analytical formula which allows the determination 
of approximate values of the magic wavelengths near resonances without
calculating 
the dynamic polarizabilities of excited states.  The optical field is
assumed to be unpolarized, though we estimate that polarization would
induce only small shifts in the magic wavelengths. 

\section{Calculations}

\subsection{{\em Ab initio} calculations}

The dynamic scalar polarizability $\alpha_a$ of atomic state $a$ is given by
(we use atomic units: $\hbar=1, m_e=1, |e|=1$)
\begin{eqnarray}
\alpha_a(\omega)&=&-\frac{1}{3(2J_a+1)}\sum_n\left[\frac{1}{E_a-E_n+\omega}
\right. \nonumber \\
&+&\left. \frac{1}{E_a-E_n-\omega}\right] \langle a||\mathbf{D}||n\rangle^2 ,
\label{eq:pol}
\end{eqnarray} 
where $J_a$ is total momentum of state $a$, $E_a$ is its energy,
$\mathbf{D}=-\sum_i \mathbf{r_i}$ is the electric dipole operator. 
Summation goes over complete set of excited states $n$.

We use the relativistic configuration interaction (CI) technique described in
our previous papers~\cite{Dzuba08,Dzuba08a,Dzuba10} to perform the
calculations. The single-electron and many-electron basis sets, the
fitting parameters and other details of present calculations are
exactly the same as in Ref.~\cite{Dzuba10}. This simple method
provides a good accuracy for low lying states of a 
many-electron atom.  However, it does not allow for the saturation of
the summation 
in Eq.~(\ref{eq:pol}) over a complete set of many-electron states. On the
other hand, the contribution of the higher-lying states in the dynamic
polarizability does not depend on frequency at small
frequencies. Therefore, for small frequencies we can rewrite
Eq.~(\ref{eq:pol}) as 
\begin{eqnarray}
\alpha_a(\omega)&=&\tilde
\alpha_a-\frac{1}{3(2J_a+1)}\sum_{n'}\left[\frac{1}{E_a-E_{n'}+\omega}  
\right. \nonumber \\
&+&\left. \frac{1}{E_a-E_{n'}-\omega}\right] \langle
a||\mathbf{D}||n'\rangle^2 , 
\label{eq:pol1}
\end{eqnarray} 
where the summation is over a limited number of low-lying near-resonant
states and a constant $\tilde \alpha_a$ is chosen in such a way that
Eq.~(\ref{eq:pol1}) at $\omega=0$ provides the correct value of the
polarizability. 

Dysprosium ground state static polarizability is known to be 166
$a_B^3$~\cite{CRC}. Static 
polarizabilities of excited states are not known and need to be
calculated. We use an approximate approach in which the dysprosium atom is
treated as a closed-shell system and the effect of electron
vacancies in the open shells is taken into account via fractional
occupation numbers. The static polarizability of a closed-shell system is
given by
\begin{equation}
\alpha_a(0) = -\frac{2}{3}\sum_{cn}\frac{\langle
  c||\mathbf{D}||n\rangle^2}{\epsilon_c-\epsilon_n}, 
\label{eq:pol2}
\end{equation} 
where the summation is over a complete set of single-electron states
including states in the core ($c$) and states above the core ($n$).
Electric dipole matrix elements are calculated using relativistic
Hartree-Fock and Hartree-Fock in external field approximations
\cite{DFSS87}. Note that core polarization needs to be included only in
one of two electric dipole matrix elements in (\ref{eq:pol2}) (see,
e.g., Ref.~\cite{DFGK02} for details).
 
We use the standard B-spline technique~\cite{Bspline} to generate a
complete set of single-electron states. An additional term is included
into the Hartree-Fock Hamiltonian to simulate the effect of
correlations. This term has the form 
\begin{equation}
\delta V(r) = - \frac{d}{2(r_0^4+r^4)},
\label{eq:dv}
\end{equation}
where $r_0$ is a cut-off parameter (we use $r_0=1$ $a_B$) and $d$
is dipole polarizability of the core. We treat $c$ as a fitting
parameter and chose it to fit the known polarizability of dysprosium's
ground state (166$a_0^3$~\cite{CRC}), which results in $d=3.7$
$a_B^3$. 

Then we perform similar calculations for the excited states of the
$4f^96s^25d$ configuration, resulting in a calculated value of the static
polarizability of 114 $a_B^3$. Note that this approach does not
distinguish between different states of the same configuration. Therefore,
static polarizabilities of all these states are assumed to be equal. This is
only true for the static polarizabilities. Dynamic polarizabilities
are different for different states due to contributions of the near
resonant states in Eq.~(\ref{eq:pol1}).

In the present paper we consider dynamic polarizabilities of four
states of dysprosium: the even ground state (GS) and three odd
long-lived excited states. The first excited state is $^7$H$^o_8$ at
$\lambda = 1322$ nm ($E$=7565.60~cm$^{-1}$), and we denote it as O1
for reference.  This state is in the telecommunications band, and
could be used for hybrid atom-photon telecom quantum information
networks.  The second excited state is the $^7$I$^o_9$ state at 
1001 nm (9990.95~cm$^{-1}$), we denote as O2.  InAs quantum dots (QDs)
emit in this wavelength range, allowing the possibility for hybrid
quantum circuits of QD single photon emitters coupled to neutral
atom-based long-lived quantum memory. O3 is the $^5$K$^o_9$ state at
741 nm (13495.92~cm$^{-1}$), which is a closed cycling transition with
a linewidth~\cite{Lu2010b} optimal for creating a narrow-line MOT.
States O2 and O3 could also be useful for resolved-sideband cooling,
as discussed below.   

We calculate dynamic polarizabilities using the Eq.~(\ref{eq:pol}) in
which we substitute transition amplitudes found from the CI
calculations~\cite{Dzuba10} and experimental energies.  We use
theoretical values in the few cases where experimental energies are
not available. Tables~\ref{t:amp0} and \ref{t:amp1} show calculated
electric dipole transition amplitudes (reduced matrix elements) used
in the calculations. The data from Table \ref{t:amp1} can be used to
calculate the lifetimes of the three excited states. The results are
5.2~ms for the O1, 2.7~ms for O2, and 21~$\mu$s for O3, though O3 has
recently been measured to be 89.3 $\mu$s~\cite{Lu2010b}.  

\begin{table}
\caption{Electric dipole transition amplitudes (reduced matrix
  elements in atomic units) used for calculation of the dynamic
  polarizability of the Dy ground state $^5$I$_8$.}
\label{t:amp0}
\begin{ruledtabular}
 \begin{tabular}{l c r c}

\multicolumn{2}{c}{State $n$} &
\multicolumn{1}{c}{$E_n$} &
\multicolumn{1}{c|}{$|A_{na}|$\footnotemark[1]} \\
&&\multicolumn{1}{c}{(cm$^{-1}$)}&
\multicolumn{1}{c}{(a.u.)}\\

\hline
$4f^{9}5d6s^2$ & $^7$H$^o_8$ &  7565 &  0.061 \\
$4f^{9}5d6s^2$ & $^7$H$^o_7$ &  8519 &  0.124 \\
$4f^{9}5d6s^2$ & $^7$I$^o_9$ &  9990 &  0.059 \\
$4f^{9}5d6s^2$ & $^7$I$^o_8$ & 12007 &  0.573 \\
$4f^{9}5d6s^2$ & $^7$G$^o_7$ & 12655 &  0.108 \\
$4f^{9}5d6s^2$ & $^5$K$^o_9$ & 13495 &  0.424 \\
$4f^{9}5d6s^2$ & $^7$I$^o_7$ & 14367 &  0.475 \\
$4f^{9}5d6s^2$ & $^5$I$^o_8$ & 14625 &  1.828 \\
$4f^{9}5d6s^2$ & $^5$H$^o_7$ & 15194 &  1.452 \\
$4f^{10}6s6p$ & $(8,0)^o_8$ & 15567 &  0.464 \\
$4f^{10}6s6p$ & $(8,1)^o_9$ & 15972 &  1.365 \\
$4f^{9}5d6s^2$ & $^7$K$^o_8$ & 16288 &  0.182 \\
$4f^{10}6s6p$ & $(8,1)^o_7$ & 16693 &  1.842 \\
$4f^{10}6s6p$ & $(8,1)^o_8$ & 16733 &  0.633 \\
$4f^{9}5d6s^2$ & $^7$K$^o_9$ & 16717 &  0.415 \\
$4f^{9}5d6s^2$ & $^7$K$^o_7$ & 17687 &  0.763 \\
$4f^{10}6s6p$ & $(8,2)^o_9$ & 17727 &  0.897 \\
$4f^{10}6s6p$ & $(8,2)^o_8$ & 18021 &  0.684 \\
$4f^{10}6s6p$ & $(8,2)^o_7$ & 18433 &  0.636 \\
$4f^{9}5d^26s$ & $^9$G$^o_7$ & 18528 &  0.067 \\
$4f^{9}5d^26s$ & $^7$H$^o_9$ & 19557 &  0.036 \\
$4f^{9}5d6s^2$ & $^5$K$^o_8$ & 19688 &  0.627 \\
$4f^{9}5d^26s$ & $^7$G$^o_9$ & 21540 &  0.523 \\
$4f^{10}6s6p$ & $(7,2)^o_9$ & 21838 &  0.513 \\
$4f^{9}5d^26s$ &    $?^o_9$ & 23271 &  0.003 \\
$4f^{10}6s6p$ & $(8,1)^o_9$ & 23737 & 12.277 \\

\end{tabular}
\footnotetext[1]{$A_{na} \equiv \langle n||\mathbf{D}|| a \rangle$.}
\end{ruledtabular}
\end{table}

\begin{table}
\caption{Electric dipole transition amplitudes (reduced matrix
  elements in atomic units) used for calculation of the dynamic
  polarizabilities of the selected three long-lived Dy excited states.}
\label{t:amp1}
\begin{ruledtabular}
 \begin{tabular}{l c r c c c}

\multicolumn{2}{c}{State $n$} &
\multicolumn{1}{c}{$E_n$} &
\multicolumn{3}{c}{$|A_{na}|$\footnotemark[1] (a.u.)} \\
&&\multicolumn{1}{c}{(cm$^{-1}$)}&
\multicolumn{1}{c}{O1\footnotemark[2]} &
\multicolumn{1}{c}{O2\footnotemark[3]} &
\multicolumn{1}{c}{O3\footnotemark[4]} \\

\hline

$4f^{10}6s^2$ & $^5$I$_8$    &     0 &  0.061 & 0.059 & 0.424 \\
$4f^{10}6s^2$ & $^5$I$_7$    &  4134 &  0.007 &       &       \\
$4f^{10}5d6s$ & $^3[8]_9$    & 17515 &  0.033 & 0.154 & 0.139 \\
$4f^{10}5d6s$ & $^3[7]_8$    & 17613 &  0.195 & 0.321 & 0.046 \\
$4f^{10}5d6s$ & $^3[6]_7$    & 18095 &  0.170 &       &       \\
$4f^{10}5d6s$ & $^3[9]_{10}$ & 18463 &        & 0.060 & 0.210 \\
$4f^{10}5d6s$ & $^3[8]_8$    & 18903 &  0.300 & 0.401 & 0.268 \\
$4f^{10}5d6s$ & $^3[7]_7$    & 18938 &  0.259 &       &       \\
$4f^{10}6s^2$ & $^3$K2$_8$   & 19019 &  0.083 & 0.093 & 0.064 \\
$4f^{10}5d6s$ & $^3[9]_9$    & 19241 &  0.079 & 0.119 & 0.015 \\
$4f^{10}5d6s$ & $^3[10]_{10}$ & 19798 &        & 0.049 & 0.192 \\
$4f^{10}5d6s$ & $^3[9]_8$    & 20194 &  0.358 & 0.336 & 0.288 \\
$4f^{10}5d6s$ & $^3[10]_9$   & 20209 &  0.041 & 0.062 & 0.010 \\
$4f^96s^26p$ &$(\frac{15}{2},\frac{1}{2})_7$ & 20614 &  2.164 &       &       \\
$4f^96s^26p$ &$(\frac{15}{2},\frac{1}{2})_8$ & 20790 &  3.737 & 4.254 & 2.911 \\
$4f^{10}5d6s$ & $^3[8]_7$    & 21074 &  0.550 &       &       \\
$4f^{10}5d6s$ & $^3[7]_8$    & 21603 &  0.362 & 0.450 & 0.130 \\
$4f^{10}5d6s$ & $^3[6]_7$    & 21778 &  0.463 &       &       \\
$4f^{10}5d6s$ & $^1[9]_9$    & 22046 &  0.079 & 0.081 & 0.116 \\
$4f^{10}5d6s$ & $^1[9]_{10}$ & 22487 &        & 0.004 & 0.112 \\
$4f^95d6s6p$ & $?_9$        & 23219 &  1.313 & 3.879 & 0.400 \\
$4f^{10}5d6s$ & $?_7$       & 23361 &  0.172 &       &       \\

\end{tabular}
\footnotetext[1]{$A_{na} \equiv \langle n||\mathbf{D}|| a \rangle$;}
\footnotetext[2]{State $a=4f^{9}5d6s^2$, \ $^7$H$^o_8, \ \ E=7565.60$ cm$^{-1}$; \ \ $\lambda = 1322$ nm;} 
\footnotetext[3]{State $a=4f^{9}5d6s^2$, \ $^7$I$^o_9, \ \ \ E=9990.95$ cm$^{-1}$; \ \ $\lambda = 1001$ nm;} 
\footnotetext[4]{State $a=4f^{9}5d6s^2$, \ $^5$K$^o_9, \ \ E=13495.92$ cm$^{-1}$; \ \ $\lambda = 741$ nm.} 
\end{ruledtabular}
\end{table}

\begin{figure}[!hbt]
\centering
\epsfig{figure=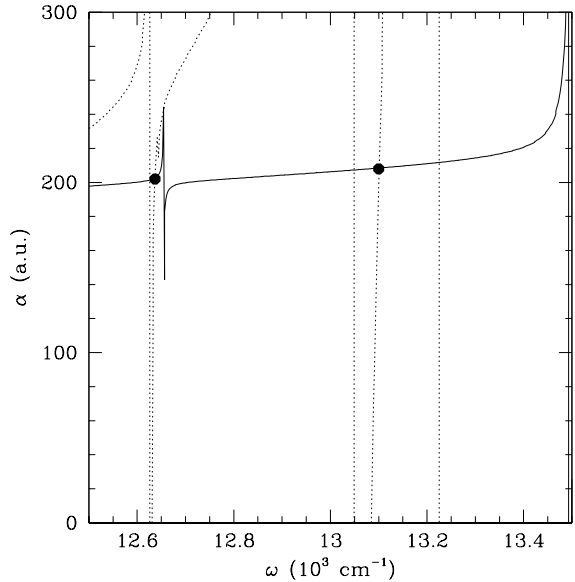,scale=0.4}
\caption{Dynamic polarizability $\alpha$ of the ground state of Dy
  (solid line) and O1 (dotted
  line) between laser frequencies (wavelengths) 12500~cm$^{-1}$ (800
  nm) and 13500~cm$^{-1}$ (741 nm). 
  Lines cross at magic frequencies, and large dots correspond to
  the most useful.}
\label{Fig:wm1}
\end{figure}

\begin{figure}[!hbt]
\centering
\epsfig{figure=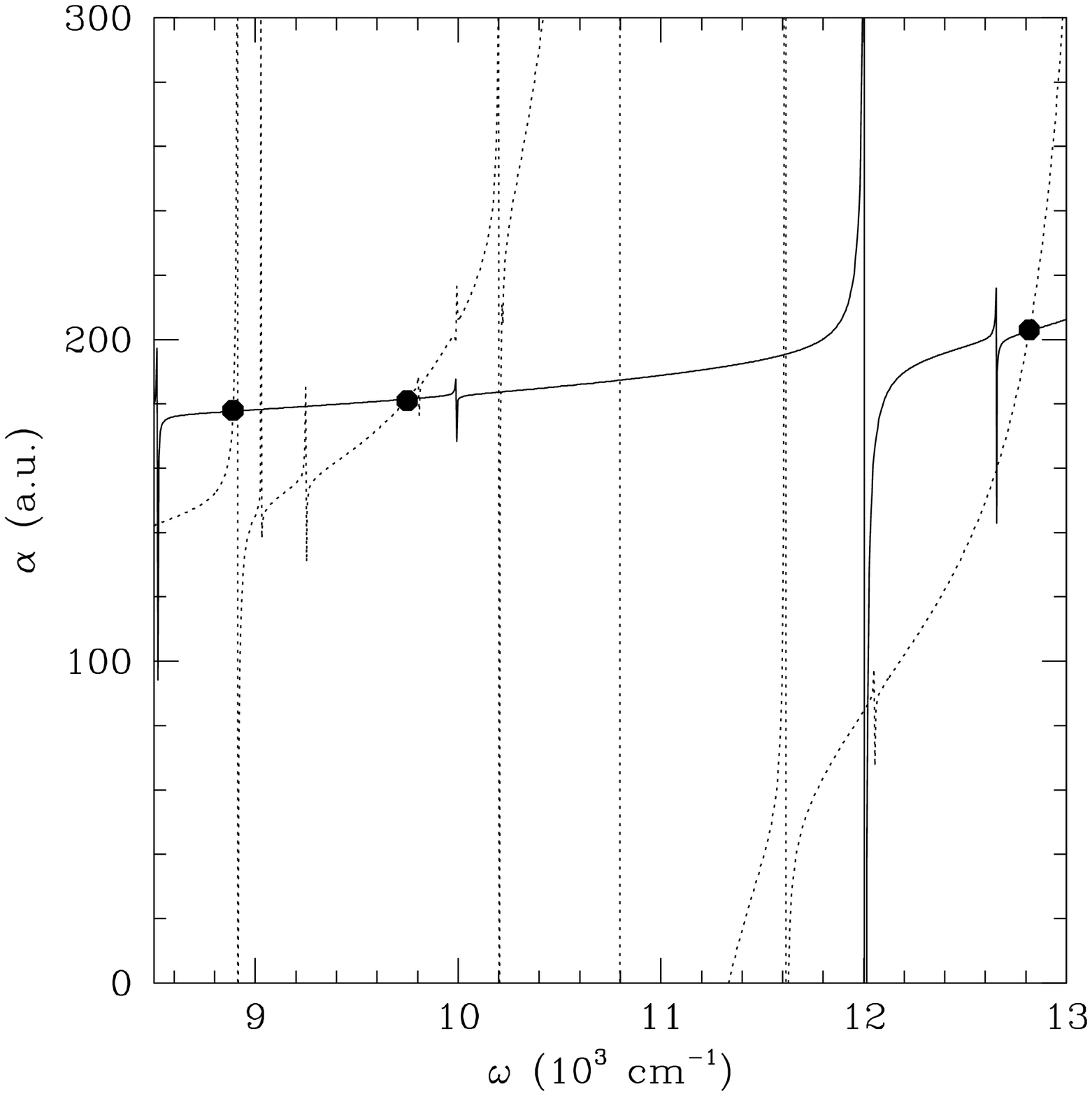,scale=0.4}
\caption{Dynamic polarizability $\alpha$ of the ground state of Dy
  (solid line) and O2 (dotted
  line) between laser frequencies (wavelengths) 8500~cm$^{-1}$ (1.18 $\mu$m) and 13000~cm$^{-1}$ (769 nm).
  Lines cross at magic frequencies, and large dots correspond to
  the most useful.}
\label{Fig:wm2}
\end{figure}

\begin{figure}[!hbt]
\centering
\epsfig{figure=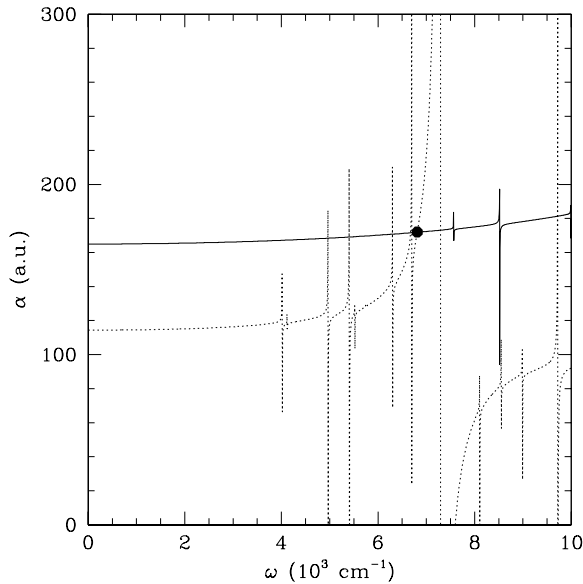,scale=0.4}
\caption{Dynamic polarizability $\alpha$ of the ground state of Dy
  (solid line) and O3 (dotted
  line).  Lines cross at magic frequencies, and the large dot corresponds to
  the most useful.}
\label{Fig:wm3}
\end{figure}

Figures \ref{Fig:wm1}, \ref{Fig:wm2} and \ref{Fig:wm3} show dynamic
polarizabilities of three pairs of states: GS and O1
(Fig. \ref{Fig:wm1}), GS and O2 (Fig. \ref{Fig:wm2}) and GS and O3
(Fig. \ref{Fig:wm3}). Lines crossing means that energy shifts of two 
states in a laser field are identical, and atoms have the same
oscillation frequency in optical dipole traps at this wavelength
regardless of whether they are in their ground or excited state.
These so-called {\em magic} wavelengths occur most often very close to
narrow resonances.

\subsection{Simple estimations}

In this subsection we present a way of estimating magic wavelengths
for complex atoms in the vicinity of narrow resonances. Although, all
magic wavelengths presented in this work are found by the many-body
calculations, the formulas of this subsection can be used to find
more magic wavelengths for dysprosium or to find magic wavelengths for
other complex atoms. We demonstrate that many-body calculations and
simple estimations give close results.

In the case of a narrow resonance, energy denominator in
Eq.~(\ref{eq:pol}) is close to zero.  
This makes it possible to write an 
approximate formula for the magic frequency corresponding to this
wavelength. Starting from the condition 
\begin{equation}
  \alpha_{GS}(\omega^*) = \alpha_{a}(\omega^*),
\label{eq:m}
\end{equation}
and presenting dynamic polarizability of the excited state in the
vicinity of the resonance $n$ in the form
\begin{eqnarray}
\alpha_a(\omega)&=&\alpha_a(0)-\frac{1}{3(2J_a+1)}\left[\frac{1}{E_a-E_n+\omega}
\right. \nonumber \\
&+&\left. \frac{1}{E_a-E_n-\omega}\right] \langle a||\mathbf{D}||n\rangle^2 ,
\label{eq:poln}
\end{eqnarray} 
we arrive at the following expression:
\begin{eqnarray}
\omega^*_{an} &=& |E_a-E_n| + \frac{E_a-E_n}{|E_a-E_n|}
\delta_{n},  \label{eq:magic} \\
\delta_{n} &=&\frac{1}{3(2J_a+1)}\frac{\langle
  a||\mathbf D|| n \rangle^2}{(\alpha_{GS}(\omega_n)-\alpha_{a}(0))}, \nonumber
\end{eqnarray}
where $\omega_n = |E_a-E_n|$ is the resonance frequency, $n$ is the
resonance number, $\langle a||\mathbf D|| n \rangle$ is the electric
dipole transition amplitude from the excited state $a$ to the
resonance state $n$, and $\alpha_{GS}(\omega_n)$ is the scalar
polarizability of the ground  state at $\omega_n$.  Note that
$\omega_n$ is the resonance frequency for the upper state, and the
polarizability of the ground state usually changes very little in the
vicinity of the $\omega_n$'s. 

In case of two closely spaced resonances in the upper state
polarizability, magic frequencies in the vicinity of resonance
energies $E_1$ and $E_2$  can be found using approximate formula
\begin{equation}
  \omega^*_{a12} = \frac{E_1+E_2}{2}+\delta_{12},
\label{eq:w2}
\end{equation}
where
\begin{equation}
  \delta_{12}=\Delta_{12}\frac{C_J\Delta_{12}\left[\alpha^{(0)}_{GS}(\omega_{12})
      - \alpha^{(0)}_a(0)\right] + A^2_2 - A^2_1}{A^2_2 + A^2_1}.
\label{eq:w2d}
\end{equation}
Here, $\Delta_{12}=(E_1-E_2)/2$, $C_J=3(2J_a+1)$, $\omega_{12}=(E_1+E_2)/2$, $A_1 =
\langle a||\mathbf{D}||1\rangle$ and $A_2 =\langle a||\mathbf{D}||2\rangle$.

\begin{table*}
\caption{Magic wavelengths ($\lambda^{*}$), frequencies ($\omega^*$, cm$^{-1}$), and
  polarizabilities ($\alpha$) for the three transitions in Dy.}
\label{t:m}
\begin{ruledtabular}
  \begin{tabular}{c r r r r r r r}
\multicolumn{1}{c}{Transition} &
\multicolumn{1}{c}{Resonance} &
\multicolumn{4}{c}{Magic frequencies} &
\multicolumn{1}{c}{$\lambda^*$} &
\multicolumn{1}{c}{$\alpha$} \\
&\multicolumn{1}{c}{$E_n-E_a$} &
\multicolumn{1}{c}{$\delta_{n}$\footnotemark[1]} &
\multicolumn{1}{c}{$\delta_{n}$\footnotemark[3]} &
\multicolumn{1}{c}{Formula} &
\multicolumn{1}{c}{Calculations} &
\multicolumn{1}{c}{(nm)} &
\multicolumn{1}{c}{(a.u.)} \\
\hline
GS -- O1            & 11337 & 4 && 11333 & 11326 & 883 & 192 \\
$E_a$=7566~cm$^{-1}$ & 11372 & 3 && 11368 & 11366 & 880 & 192 \\
 $\lambda = 1322$ nm                   & 13224 & 537 && 12686 & 12638 & 791 & 202 \\
                    & 13136\footnotemark[2] && -35 & 13101 & 13100 & 763 & 208 \\
                    &       &&     &       &       &     &     \\

GS -- O2            & 7523 & 1 && 7522 & 7521 & 1330 & 172 \\
$E_a$=9991~cm$^{-1}$ & 7622 & 4 && 7617 & 7613 & 1314 & 174 \\
 $\lambda = 1001$ nm                     & 8912 & 7 && 8905 & 8891 & 1125 & 178 \\
                    &      &   &&       & 9749 & 1026 & 181 \\
%                    & 10202 & 5 && 10197 & 10208 & 980 & 82 \\ % ???????????
                    & 13227 & 483 && 12744  & 12817 & 780 & 203 \\
                    &       &&     &       &       &     &     \\

GS -- O3            & 5407 & 4 && 5403 & 5401 & 1850 & 163 \\
$E_a$=13496~cm$^{-1}$ & 6302 & 2 && 6300 & 6297 & 1588 & 171 \\
  $\lambda = 741$ nm                   & 6697 & 4 && 6693 & 6671 & 1494 & 172 \\
                   & 7293 &  479 &&  6814 &  6812 &  1468 & 172 \\
                   & 9722 &    8 &&  9714 &  9716 &  1029 & 181 \\

\end{tabular}
\footnotetext[1]{Using formula (\ref{eq:magic}), $\omega^*_{an}=E_n-E_a-\delta_{n}$;}
\footnotetext[2]{$E_n=(20614+20790)/2-7566$;}
\footnotetext[3]{Using formulas (\ref{eq:w2}-\ref{eq:w2d}), $\omega^*_{a12}=(E_1+E_2)/2-E_a+\delta_{12}$.}
\end{ruledtabular}
\end{table*}

One does not need to know dynamic polarizability of an excited state to
find magic frequencies using Eqs.~(\ref{eq:magic}) or
(\ref{eq:w2}). However, one still 
needs to know the dynamic polarizability of the ground state and
the relevant transition amplitudes.  An approximate solution can be found
using the following rules:
\begin{itemize}
\item The dynamic polarizability of the ground state is fitted very well
  within energy interval from 0 to 0.05 a.u.~by
\begin{eqnarray}
&  \alpha_{GS}(\omega) =
  \left(\frac{2.955}{\omega+0.10815}-\frac{2.955}{\omega-0.10815}\right)+
\nonumber \\
&    110 + 3000\omega^2.
\label{eq:fit}
\end{eqnarray}
We keep here numerical parameters for the dominant contribution to
$\alpha_{GS}$, which is due to the transition to the 421-nm state with
energy E= 23737 cm$^{-1}$ = 0.10815 a.u.  This transition dominates due
to the largest value of the transition amplitude: $\langle a||\mathbf
D|| n \rangle$ =12.277 a.u. (see Table~\ref{t:amp0}). The last two terms in
Eq.~(\ref{eq:fit}) fit the contribution of all other transitions in the
energy interval $0<\omega <0.05$ a.u. Resonances within this interval are
ignored since they are too narrow for any practical importance.
\item Transition amplitudes can be estimated using approximate
  selection rules. If the difference of the total angular momentum $L$ 
  between two states is larger than 1, or if total spin $S$ of the
  states is not equal, then the amplitude is not zero due to
  relativistic corrections but is likely to be small. One can use for a
  rough estimate $A$=0.1 a.u. A similar estimation can be used if the
  transition is suppressed by configuration mixing, i.e. the transition
  between leading configurations cannot be reduced to a
  single-electron allowed electric dipole transition. If no selection
  rules are broken, the amplitude is likely to be large and one can use
  $A$=3 a.u.~as a rough estimate.
\item Polarizability of the excited state at zero frequency can be
  estimated using Eq.~(\ref{eq:pol}) with experimental energies
  and with the amplitude estimated using the procedure in the
  previous paragraph.
\end{itemize}
This procedure can help in estimating magic wavelengths not only for
the transitions considered in present paper but also for some other
transitions. The main condition for it to work is that magic wavelength
should be close to a resonance so that resonance term dominates
in Eq.~(\ref{eq:pol}).  

\section{Results}
Table \ref{t:m} shows the magic wavelengths and corresponding
polarizabilities for the three
transitions in Dy.  We substitute calculated transition amplitudes from
Table~\ref{t:amp1} when using the analytical formulas (\ref{eq:magic})
and (\ref{eq:w2}).
The column marked as {\em calculations} presents magic frequencies
which come from numerical calculations. Magic wavelengths in the next
column correspond to calculated frequencies.
Most of the magic frequencies are due to very narrow
resonances and might be inconvenient for practical use due to optical
dipole trap frequency instabilities and enhanced spontaneous emission.
However, there are magic wavelengths for each of the three transitions
where the resonance is not very narrow or even absent. They are
$\lambda = 791$ nm ($\omega^*=12638$~cm$^{-1}$) 
and $\lambda = 763$ nm ($\omega^*=13100$~cm$^{-1}$) for the 1322-nm
GS-O1 transition; $\lambda = 1125$ nm ($\omega^*=8891$~cm$^{-1}$),
$\lambda = 1026$ nm ($\omega^*=9749$~cm$^{-1}$) 
and $\lambda = 780$ nm ($\omega^*=12817$~cm$^{-1}$) for the 1001-nm
GS-O2 transition; and $\lambda = 1029$ nm ($\omega^*=9716$~cm$^{-1}$)
and $\lambda = 1468$ nm ($\omega^*=6812$~cm$^{-1}$) for the 741-nm
GS-O3 transition. 

The magic frequency $\omega^*=13100$~cm$^{-1}$ for the GS-O1
transition is between two close resonances at $\omega=13149$~cm$^{-1}$
and $\omega^*=13224$~cm$^{-1}$ which correspond to transitions from
the $^7$H$^{\rm o}_8$ state at $E=7566$~cm$^{-1}$ to the close states
of the $4f^96s^26p_{1/2}$ configuration at  $E=20614$~cm$^{-1}$ and
$E=20789$~cm$^{-1}$. Using formulas (\ref{eq:w2}-\ref{eq:w2d}) gives
a very accurate estimate of the magic frequency (see
Table~\ref{t:m}). It is interesting to note that there is a frequency
interval for the GS-O1 transition where polarizabilities of two states
come very close to each other but don't cross: $\delta \alpha/\alpha
\le 2$\% for $12121 < \omega < 12183$ ~cm$^{-1}$.

The magic wavelengths $\lambda = 780$ and 1026 nm for the GS-O2
transition and $\lambda = 1468$ nm for GS-O3 do not correspond to any
very-close resonance, and the values of the polarizabilities coincide
by chance rather than due to a resonance. Therefore, these magic
wavelengths are the least sensitive to laser frequency fluctuations
and are most promising for resolved-sideband cooling, precision
measurement, and QIP applications.  The GS-O2 transition magic
wavelengths can be reached with high optical power using a Ti:sapphire
or tapered amplified diode laser for 780 nm and diode laser or fiber
laser for 1026 nm.  The GS-O3 transition magic wavelength at 1468 nm
could be reached with a low-power diode laser, which could perhaps be
doped-fiber amplified, and the 1029-nm wavelength could be reached
with a fiber laser.   

Future work will explore the role of laser polarization on the magic
wavelength position.  Such a calculation is beyond the scope of this
present work, but we estimate that the shifts will be small since the
magic wavelengths for unpolarized light occur in proximity to
resonances. 

Optical dipole traps at these wavelengths, far from the broad Dy
transitions, would be suitable for lattice confinement in the
Lamb-Dicke regime without undue heating.  For example, 1D lattice
confinement at the 780-nm magic wavelength with 0.5 W provides ample
trap depth with sub-1 Hz scattering rates.  With larger laser
intensities, suitable trap depths and low scattering rates can be
achieved at the other magic wavelengths.  Vibrational spacing can be
many tens of kHz, which is large enough for resolved-sideband cooling
on the 2 kHz-wide 741-nm transition~\cite{Lu2010b}.  For rapid
cooling, the 50 Hz-wide 1001-nm transition---much narrower than any
trap frequencies in a typical 3D optical lattice---would need to be
broadened via a quenching transition~\cite{Riehle07}, and the optical
dipole trap magic wavelength would need to be adjusted to compensate
the Stark shift from the quenching laser.  Resolved-sideband cooling
on these narrow transitions in a 3D optical lattice may provide an
alternative route to quantum degeneracy~\cite{Weiss02} versus
evaporative cooling, which may fail due to (as yet unmeasured)
unfavorable scattering properties in this highly dipolar gas. 

\section*{Acknowledgments}

We thank J. Ye, I. Deutsch, and N. Burdick for discussions.  The work
was funded in part by the Australian Research Council (V.A.D.,
V.V.F.), the NSF (PHY08-47469) (B.L.L.), AFOSR (FA9550-09-1-0079)
(B.L.L.), and the Army Research Office MURI award W911NF0910406
(B.L.L.).


\begin{thebibliography}{999}

\frenchspacing

\bibitem{Dzuba86} V. A. Dzuba, V. V. Flambaum, I. B. Khriplovich,
%Enhancement of P- and T-nonconserving effects in rare-earth atoms,
Z. Phys. D: {\it Atoms, Molecules and Clusters}, {\bf 1}, 243-245 (1986).

\bibitem{Dzuba94}   V. A. Dzuba, V. V. Flambaum, and M. G. Kozlov,
% PNC in Dy
Phys. Rev. A, {\bf 50}, 3812 (1994).

\bibitem{Budker94} D. Budker, D. DeMille, E. D. Commins, and M. S. Zolotorev,
    Phys. Rev. A {\bf 50}, 132 (1994).

\bibitem{Budker97} A. T. Nguyen, D. Budker, D. DeMille, and M. Zolotorev,
% Search for parity nonconservation in atomic dysprosium
Phys. Rev. A {\bf 56}, 3453 (1997).

\bibitem{Dzuba10}   V. A. Dzuba and V. V. Flambaum,
Phys. Rev. A, {\bf 81}, 052515 (2010).

\bibitem{Dzuba99a} V. A. Dzuba, V. V. Flambaum, and J. K. Webb,
% Space-Time Variation of Physical Constants and Relativistic 
% Corrections in Atoms, 
Phys. Rev. Lett., {\bf 82}, 888 (1999).

\bibitem{Dzuba99b} V. A. Dzuba, V. V. Flambaum, J. K. Webb,
% Calculations of the Relativistic Effects in Many-Electron Atoms and 
% Space-Time Variation of Fundamental Constants,
Phys. Rev. A, {\bf 59}, 230 (1999).

\bibitem{Dzuba03}  V. A. Dzuba, V. V. Flambaum, and M. V. Marchenko,
%Relativistic effects in Sr, Dy, YbII and YbIII and search for variation
%of the fine structure constant,
Phys. Rev. A, {\bf 68}, 022506 (2003).

\bibitem{Budker04} A.-T. Nguyen, D. Budker, S. K. Lamoreaux, 
and J. R. Torgerson,
%Title: Towards a sensitive search for variation of the fine-structure 
%constant using radio-frequency E1 transitions in atomic dysprosium 
%Source: PHYSICAL REVIEW A 69 (2): Art. No. 022105 FEB 2004 
Phys.  Rev.  A{\bf 69}, 022105 (2004).

\bibitem{Budker07} A. {A. Cing\"{o}z}, A. Lapierre, A.-T. Nguyen, N. Leefer, 
D. Budker, S. K. Lamoreaux, and J. R. Torgerson,
%Title: Limit on the temporal variation of the fine-structure constant 
%using atomic dysprosium 
Phys.  Rev.  Lett. {\bf 98}, 040801 (2007).

\bibitem{Dzuba08}  V. A. Dzuba and V. V. Flambaum, 
%       Relativistic corrections to transition frequencies of 
%       Ag~I, Dy~I, Ho~I, Yb~II, Yb~III, Au~I and Hg~II
%       and search for variation of the fine structure constant,
%       arXiv:0712.3621 (2007);
       Phys. Rev. A, {\bf 77}, 012515 (2008).



\bibitem{Budker07a} S. J. Ferrell, {A. Cing\"{o}z}, A. Lapierre,
  A.-T. Nguyen, N. Leefer, D. Budker, V. V. Flambaum, S. K. Lamoreaux,
  and J. R. Torgerson, 
  Phys. Rev. A {\bf 76}, 062104 (2007).


\bibitem{Lev} M. Lu, S. H. Youn, and B. L. Lev,
% arXiv:0912.0050 (2009).
%    Title: Trapping ultracold dysprosium: a highly magnetic gas for
%    dipolar physics 
Phys. Rev. Lett. {\bf 104}, 063001 (2010).

\bibitem{Budker08} N. Leefer, {A. Cing\"{o}z}, D. Budker,
  S. J. Ferrell, V. V. Yashchuk, A. Lapierre, A.-T. Nguyen, 
  S. K. Lamoreaux, and J. R. Torgerson, in {\it Proceedings of the 7th
    Symposium Frequency Standards and Metrology, Asilomar, October
    2008}, edited by Lute Maleki, World Scientific, pp. 34-43.

\bibitem{Leefer} N. Leefer, {A. Cing\"{o}z}, B. Gerber-Siff, A. Sharma,
  J. R. Torgerson, and D. Budker,
% arXiv:0912.2133 (2009).
%    Title: Transverse laser cooling of a thermal atomic beam of dysprosium
  Phys. Rev. A {\bf 81}, 043427 (2010).

\bibitem{Youn2010a} S.-H. Youn, M. Lu, U. Ray, and B. L. Lev,
  Phys. Rev. A {\bf 82}, 043425 (2010). 
 
\bibitem{Youn2010b} S.-H. Youn, M. Lu, and B. L. Lev,  Phys. Rev. A
  {\bf 82}, 043403 (2010). 

\bibitem{Berglund:2008} A. J. Berglund, J. L. Hanssen, and J. J. McClelland,
Phys. Rev. Lett. {\bf 100}, 113002 (2008).

\bibitem{Wineland89} F. Diedrich, J. C. Bergquist, W. M. Itano, and
  D. J. Wineland,  
Phys. Rev. Lett. {\bf 62}, 403 (1989).

\bibitem{Katori03} T. Ido and H. Katori, 
Phys. Rev. Lett. {\bf 91}, 053001 (2003).

\bibitem{Katori08} J. Ye, H. J. Kimble, and H. Katori,
Science {\bf 320}, 1734 (2008).

\bibitem{Riehle07} Ch. Grain, T. Nazarova, C. Degenhardt, F. Vogt,
  Ch. Lisdat, E. Tiemann, U. Sterr, and F. Riehle,  
Eur. Phys. J. D. {\bf 42}, 317 (2007).

\bibitem{Dzuba08a}  V. A. Dzuba and V. V. Flambaum,
       Phys. Rev. A, {\bf 77}, 012514 (2008).

\bibitem{CRC}T. M. Miller, in {\em Handbook of Chemistry and Physics},
Ed. D. R. Lide (CRC, Boca Raton 2000).

\bibitem{DFSS87} V. A. Dzuba, V. V. Flambaum, P. G. Silvestrov, O. P. Sushkov,
%Correlation potential method for the calculation of energy 
%levels, hyperfine structure and E1 transition amplitudes in 
%atoms with one unpaired electron,
J. Phys. B: {\it At. Mol. Phys.}, 
{\bf 20}, 1399-1412 (1987).

\bibitem{DFGK02} V. A. Dzuba, V. V. Flambaum, J. S. M. Ginges, and M. G. Kozlov,
%Electric dipole moments of Hg, Xe, Rn, Ra, Pu, and TlF induced by
%nuclear Schiff moment and limits on time-reversal violating interactions.
Phys. Rev. A, {\bf 66}, 012111 (2002).

\bibitem{Bspline} W. R. Johnson, and J. Sapirstein, 
      Phys. Rev. Lett. {\bf 57}, 1126 (1986).

\bibitem{Lu2010b} M. Lu, S.-H. Youn, and B. L. Lev,  arXiv:1009.2962 (2010).

\bibitem{Weiss02} M. Olshanii and D. Weiss, 
Phys. Rev. Lett. {\bf 89}, 090404 (2002).

\end{thebibliography}
\end{document}